\documentclass[10pt,draft,showpacs,amsmath,amsfonts,amssymb,eqsecnum,twocolumn]{revtex4}
\usepackage[T1]{fontenc}
\usepackage{color}
\usepackage{graphicx}
\usepackage{pstricks}

\usepackage{graphicx}
 \usepackage{bm}
\def\a{\alpha}
\def\b{\beta}
\def\g{\gamma}
\def\d{\delta}

\def\vp{\varphi}

\def\a{\alpha}
\def\b{\beta}

\def\g{\gamma}
\def\d{\delta}
\def\g{\gamma}

\def\sint\rightarrow\!\!\!\!\!\!\!\!{\int}
 
\begin{document}


 \title{Finsler-type modification of the Coulomb law }
 \author{ Yakov Itin}
 \email{ itin@math.huji.ac.il}
\affiliation{Institute of Mathematics, The Hebrew University of
  Jerusalem \\ and Jerusalem College of Technology, Jerusalem,
  Israel}
 \author{Claus L\"{a}mmerzahl}
 \email { claus.laemmerzahl@zarm.uni-bremen.de}
\affiliation{ZARM, University of Bremen, Am Fallturm, 28359 Bremen, Germany}
 \author{Volker Perlick}
 \email { volker.perlick@zarm.uni-bremen.de}
\affiliation{ZARM, University of Bremen, Am Fallturm, 28359 Bremen, Germany}



 \begin{abstract}
\noindent
Finsler geometry is a natural generalization of pseudo-Riemannian geometry. 
It can be motivated e.g. by a modified version of the Ehlers-Pirani-Schild 
axiomatic approach to space-time theory. Also, some scenarios of quantum 
gravity suggest a modified dispersion relation which could be phrased in 
terms of Finsler geometry. On a Finslerian spacetime, the Universality of 
Free Fall is still satisfied but Local Lorentz Invariance is violated in a way not
covered by standard Lorentz Invariance Violation schemes. In this paper we 
consider a Finslerian modification of Maxwell's equations. The corrections to 
the Coulomb potential and to the hydrogen energy levels are computed. We 
find that the Finsler metric corrections yield a splitting of the energy levels.
Experimental data provide bounds for the Finsler parameters.
\end{abstract}
\maketitle

\section{Introduction}\label{sec:Intro}

\noindent
A widely expected consequence of a (still-to-be-found) theory 
of quantum gravity is a small modification of General Relativity. 
Such a modification may be encoded in a scalar--tensor theory as 
it comes out from the low energy limit of string theory leading 
e.g. to a violation of the Universality of Free 
Fall~\cite{Damouretal2002,Damouretal2002a}. 
Other consequences might be that, in addition to the metric, there 
could be a further geometric field like torsion leading to an 
effective Riemann--Cartan geometry. 

Another modification of the usual peudo-Riemannian geometry 
is Finsler geometry. It comes about naturally in some scenarios 
inspired from quantum gravity. E.g., it was shown 
in~\cite{GirelliLiberatiSindoni2007} that a modified dispersion 
relation suggested by quantum gravity can be interpreted in 
terms of Finsler geometry. Further motivation comes from Very 
Special Relativity~\cite{CohenGlashow06}: As demonstrated 
in~\cite{GibbonsGomisPope07}, some deformations of Very Special
Relativity lead in a natural way to Finsler geometry. Finsler 
geometry has also been considered in the context of Analogue
Gravity~\cite{BarceloLiberatiVisser11}.

Finsler geometry is a framework which still respects the Universality 
of Free Fall but violates Local Lorentz Invariance. The way in which 
Local Lorentz Invariance is violated is beyond usual Lorentz Invariance 
Violation schemes like the $\chi-g$ formalism~\cite{Ni77}, the 
$TH\epsilon\mu$ framework~\cite{ThorneLeeLightman73} or the Standard 
Model Extension~\cite{KosteleckyMewes02}. Furthermore, though the 
Universality of Free Fall is valid in a Finslerian setting, gravity 
cannot be transformed away locally~\cite{Lammerzahl2011}, that is, there 
is no Einstein elevator. On a more basic level, a Finslerian geometry 
may result from a relaxed version of the Ehlers-Pirani--Schild 
axiomatics~\cite{EPS72} by not requiring  the world--function to be 
twice differentiable. 

Therefore, in view of considering all possible deviations from standard Riemannian geometry reflecting effects from quantum gravity, and in view of more fundamental issues, it might be of general interest to study further consequences of Finsler geometry. Since electromagnetic phenomena provide very precise tools for exploring the geometry of space--time, in this paper we will set up a generalization of Maxwell's equations in a Finslerian space--time and derive possible consequences for atomic physics which can be compared with experiments.

\section{Finsler geometry}\label{sec:Finsler}
\subsection{Positive definite Finsler structures}
\noindent
The central idea of Finsler geometry was already proposed by Riemann in his famous 
habilitation lecture devoted to the geometry of curved manifolds \cite{Riem}. In 
parallel to the (Riemannian) geometry based on a second rank symmetric non-degenerate 
metrical tensor $g_{\alpha \beta}(x)$ with the line element $ds^2 = 
g_{\alpha \beta}(x) dx^{\alpha} dx^{\beta}$, Riemann briefly discussed a geometry 
based on a fourth-rank totally symmetric tensor $g_{\alpha \beta \gamma \delta}(x)$ 
with the line element 
\begin{equation}\label{2-FRie}
ds^4=g_{\alpha \beta \gamma \delta}(x)dx^{\alpha}dx^{\beta}dx^{\gamma}dx^{\delta}\,.
\end{equation}
An intensive study and a further generalization of this type of geometry was given by Finsler \cite{Fins} in 1918 in his Dissertation. 
Finsler geometry is based on a \emph{Finsler function} $F(x, y)$ that assigns a length
\begin{equation}\label{eq:length}
S = \int _{s_1}^{s_2} F \big( x (s), \dot{x}(s) \big) \, ds
\end{equation}
to each curve. 
One requires that $F(x, y )$ is positively homogeneous of degree one,
\begin{equation}\label{2-F}
F(x, \lambda y ) = \lambda F(x, y) \quad \mathrm{for} \: \, \lambda >0 \, ,
\end{equation}
to make sure that the length of a curve is independent of its parametrization, and that the 
\emph{Finsler metric}
\begin{equation}\label{Hess}
g_{\alpha \beta}(x,y)=
\frac{\partial^2 \big( F(x,y)^2 \big)}{\partial y^{\alpha}\partial y^{\beta}}
\end{equation}
is positive definite for all $y \neq 0$. 

The unparametrized geodesics of a Finsler geometry are the extremals of the 
length functional (\ref{eq:length}) where the endpoints are kept fixed. 
The 
affinely parametrized geodesics are the extremals of the ``energy functional''
\begin{equation}\label{eq:energy}
E = \int _{s_1}^{s_2} F \big( x (s), \dot{x}(s) \big)^2 \, ds
\end{equation}
where the endpoints and the parameter interval are kept fixed. Riemannian geometry
is, of course, a special case of Finsler geometry, characterized by the additional
property that the metric $g_{\alpha \beta}$ is independent of $y$.

The theory of positive definite Finsler metrics, which is detailed e.g. in 
~\cite{Rund} and \cite{BCS}, has several applications to physics, where
the underlying manifold is to be interpreted as three-dimensional space, so 
the greek indices take values 1,2,3. E.g., the Lagrangian of a charged particle 
in a magnetostatic field is given by a Finsler 
function of the Randers form
\begin{equation}\label{Randers}
F(x,y) =\sqrt{h_{\mu \nu}(x)y^{\mu}y^{\nu}}+A_{\mu}(x)y^{\mu} 
\end{equation}
where $h_{\mu \nu} (x)$ is a Riemannian metric (i.e., positive definite)
and $A_{\mu}(x)$ is a one-form. It can be shown that the corresponding 
Finsler metric (\ref{Hess}) is, indeed, positive definite for all $y \neq 0$ 
provided that $F(x,y) > 0$ for all $y \neq 0$, see \cite{BCS}, Section 11.1.  
To mention another example, light propagation in an anisotropic medium that
is time-independent is characterized by two positive definite spatial 
Finsler metrics \cite{Perlick1,Perlick2}. If these two metrics
coincide (i.e., if there is no birefringence), they are necessarily Riemannian 
\cite{HehlLammerzahl2004,Itin:2005iv}. Positive definite Finsler metrics have
also been used for describing the propagation of seismic waves, see 
e.g. \cite{Cerveny2002}. 
 
\subsection{Finsler structures of Lorentzian signature}
In applications to space--time physics, the Euclidean signature of the metric must be 
replaced by a Lorentzian signature. Following Beem \cite{Be70}, this can be done by 
considering, instead of the function $F(x,y)^2$, a Lagrangian $L(x,y)$ that may take 
positive, zero and negative values. (Notice that it is the \emph{square} of the Finsler function 
that enters into the definition of the metric tensor (\ref{Hess}).)

More precisely, a Finsler structure of Lorentzian signature is a function 
$L(x,y)$ that is positively homogeneous of degree two,
\begin{equation}\label{eq:homL}
L(x, \lambda y ) = \lambda ^2 L(x, y) \quad \mathrm{for} \: \, \lambda >0 \, ,
\end{equation}
and for which the Finsler metric
\begin{equation}\label{eq:g}
g_{ij}(x,y)= \frac{\partial^2 L(x,y)}{\partial y^i\partial y^j}
\end{equation}
is non-degenerate and of Lorentzian signature for all $y \neq 0$. (Actually, it
is recommendable to relax the latter condition by requiring the conditions on the Finsler 
metric to hold only for \emph{almost} all $y \neq 0$, see \cite{LaemmerzahlPerlickHasse2012}.) 
In applications to physics, the underlying manifold is to be interpreted as space--time, so 
the latin indices take values 0,1,2,3.

The homogeneity condition (\ref{eq:homL}) implies that 
\begin{equation}\label{Lg}
L(x, y ) = \dfrac{1}{2} \, g_{ij}(x,y) y^iy^j \, .
\end{equation}
The affinely parametrized geodesics of such a Finsler structure 
are, by definition, the extremals of the ``energy functional''  
\begin{equation}\label{eq:energy2}
E = \int _{s_1}^{s_2} L \big( x (s), \dot{x}(s) \big) \, ds \, . 
\end{equation}
The homogeneity condition assures that $L$ is a constant of motion, so the geodesics can be classified as
timelike ($L<0$), lightlike ($L=0$) and spacelike ($L>0$).

\section
{\bf Maxwell's equations on a flat Finsler space--time}\label{sec:Maxwell}
\noindent
In this section we discuss how Maxwell's equations must be modified if the underlying 
space--time is Finslerian. We mention that there are different views on this 
issue, see e.g. Pfeifer and Wohlfarth \cite{Pfeifer:2011tk} for an alternative approach.
We follow a line of thought that was sketched already in the appendix of 
\cite{LaemmerzahlPerlickHasse2012}. Our guiding principles are that the electromagnetic
field strength should be a field on space--time (and not on the tangent bundle, as in
\cite{Pfeifer:2011tk}), and that the lightlike Finsler geodesics should be the
bicharacteristics (i.e., the ``rays'') of Maxwell's equations.

\subsection{Flat Finsler space-times}
As in this paper we are interested in laboratory experiments, where space--time
curvature plays no role, we assume that the underlying Finsler structure is flat.
We prescribe this Finsler structure in terms of a Lagrangian, following Beem's 
definition. The flatness assumption means that we can choose the coordinates 
such that the Lagrangian is independent of $x$,
\begin{equation}\label{eq:L}
L(y)\, = \, \dfrac{1}{2} \, g_{ij}(y)y^iy^j \, .
\end{equation}
This is analogous to the pseudo-Riemannian case where the flatness assumption 
means that the coordinates can be chosen such that the $g_{ij}$ are independent 
of $x$. Here and in the following, latin indices take values 0,1,2,3 and greek indices
take values 1,2,3.

As a consequence of (\ref{eq:homL}) and (\ref{eq:g}), the Finsler metric 
is homogeneous of degree zero,
\begin{equation}\label{eq:homg}
y^k \dfrac{\partial g_{ij}( y)}{\partial y^k} \, = \, 0 \, ,
\end{equation}
and its derivative is totally symmetric,
\begin{equation}\label{eq:cycg}
\dfrac{\partial g_{ij}( y)}{\partial y^k} \, = \, 
\dfrac{\partial g_{ki}( y)}{\partial y^j} \, = \, 
\dfrac{\partial g_{jk}( y)}{\partial y^i} \, .
\end{equation}

We will later assume that $g_{ij}(y)$ is a small perturbation of 
the Minkowski metric, but in this section we will not need this 
specification.

\subsection{Hamiltonian vs Lagrangian formalism}
Recall that the lightlike geodesics of our Finsler structure are 
the extremals of the functional (\ref{eq:energy}) with $L(x,y)=0$. 
In the case at hand, where $L$ is assumed to be independent of 
$x$, the lightlike geodesics are the straight lines $x^i(s)
= a^i +y^i s$ with $L(y)=0$. 
To characterize these curves in terms of a Hamiltonian,
rather than in terms of a Lagrangian, we introduce the canonical momenta
\begin{equation}\label{eq:p}
p_i = \dfrac{\partial L(y)}{\partial y^i}
\end{equation}
and the Hamiltonian 
\begin{equation}\label{eq:H}
H(p) = p_i y^i - L(y) \, .
\end{equation}
In (\ref{eq:H}), the $y^i$ must be expressed in terms of the 
$p_j$ with the help of (\ref{eq:p}). The non-degeneracy of the 
Finsler metric guarantees that this can be
done for all $y \neq 0$. 

With (\ref{eq:L}), (\ref{eq:cycg}) and (\ref{eq:homg}) we see that (\ref{eq:p})
can be written more explicitly as
\begin{equation}\label{eq:pLy}
p_i = g_{in}(y)y^n +
\frac 12\dfrac{\partial g_{mn}(y)}{\partial y^i}y^my^n
= g_{in}(y)y^n \, .
\end{equation}
Thereupon, the Hamiltonian (\ref{eq:H}) reads
\begin{equation}\label{eq:H-x}
H(p) = \frac 12 g^{ij}(p) p_ip_j
\end{equation}
where 
\begin{equation}\label{eq:ginv}
g^{ij} (p) \, = \, 
\dfrac{\partial ^2 H(p)}{\partial p_i \partial p_j}
\end{equation}
is the inverse of $g_{jk} (y)$, with the $y^i$
expressed in terms of the $p_i$ by (\ref{eq:p}). 
In accordance with (\ref{eq:homg}) and (\ref{eq:cycg}) we have
\begin{equation}\label{eq:homginv}
p_k \dfrac{\partial g^{ij} (p)}{\partial p_k} \, = \, 0 \, ,
\end{equation}
\begin{equation}\label{eq:cycginv}
\dfrac{\partial g^{ij} (p)}{\partial p_k} \, = \, 
\dfrac{\partial g^{ki} (p)}{\partial p_j} \, = \, 
\dfrac{\partial g^{jk} (p)}{\partial p_i} \, .
\end{equation}
The Hamiltonian $H$ is homogeneous of degree two with respect to $p$, i.e. 
\begin{equation}\label{eq:homH}
p_kH^k(p) = 2H(p)
\end{equation}
where we have introduced, as an abbreviation, 
\begin{equation}\label{eq:Hi}
H^k(p) \, = \, \dfrac{\partial H (p) }{\partial p_k} \, = \,
g^{kj}(p) p_j
 \, .
\end{equation}
The lightlike Finsler geodesics (i.e., the lightlike straight lines in the case at hand) are
the solutions to Hamilton's equations with $H(p)=0$.

\subsection{Modified Maxwell's equations}
If the space-time metric is the unperturbed Minkowski metric, 
$g^{jk} = \eta ^{jk}$ where $\big( \eta ^{jk} \big) = 
\mathrm{diag} (-1,1,1,1)$, Maxwell's equations read
\begin{equation}\label{eq:Max1}
\partial _l F_{jk} + \partial _j F_{kl} + \partial _k F_{lj} \, = \, 0 \, .
\end{equation}
\begin{equation}\label{eq:Max2a}
\eta ^{kl} \partial _l F_{kj} = - \mu _0 J_j \, .
\end{equation}
Here the two-form $F_{kj}$ is the electromagnetic field strength, 
$J_j$ is the current density and $\mu _0$ is the permeability of 
the vacuum. If the current is given, (\ref{eq:Max1}) and 
(\ref{eq:Max2a}) give  a system of first-order partial 
differential equations for the electromagnetic field 
strength. 

If we replace the Minkowski metric $\eta ^{kl}$ with our flat Finsler 
metric $g^{lk}(p)$, we see that there is no reason to modify 
(\ref{eq:Max1}) because it does not involve the metric. As to 
(\ref{eq:Max2a}), it is most natural to replace
\begin{equation}\label{eq:pdo}
\eta ^{kl} \partial _l \mapsto g^{kl} ( - i \partial ) \partial _l
\end{equation}
where $i$ is the imaginary unit and $g^{kl} (- i \partial )$ 
stands for the expression that results if in $g^{kl}(p)$ the 
$p_j$ are replaced with $- i \partial _j = - i \partial / \partial x^j$.
As $g^{kl}(p)$ is not in general a polynomial in the momentum
coordinates, $g^{kl} (- i \partial ) \partial _l$ is not in general a 
differential operator but rather a pseudo-differential operator. 
(For background material on pseudo-differential operators see 
e.g.~\cite{Taylor1991}.) With the replacement (\ref{eq:pdo}),
the Maxwell equation (\ref{eq:Max2a}) becomes a pseudo-differential
equation,
\begin{equation}\label{eq:Max2g}
g^{kl} ( - i \partial ) \partial _l F_{kj} = - \mu _0 J_j \, . 
\end{equation}
By (\ref{eq:Hi}), this equation can be equivalently 
rewritten as
\begin{equation}\label{eq:Max2}
i H^k(-i \partial ) F_{kj} = - \mu _0 J_j \, . 
\end{equation}
As the current and the field strength are both real, the operator
$i H^k(-i \partial )$ should map real functions to real functions.
This is the case if the Hamiltonian is even, $H(-p) = H(p)$, i.e.,
if the homogeneity property (\ref{eq:homL}) is true also for negative 
$\lambda$.
If this condition is satisfied, (\ref{eq:Max1}) and (\ref{eq:Max2}) 
determine a perfectly reasonable dynamical system for the field strength
if the current is given. Note that  if $H$ satisfies the property 
\begin{equation}\label{eq:ihom}
H(-i p) = -H(p) \, ,
\end{equation}
we may write 
\begin{equation}\label{eq:ihom2}
i H^k(-i \partial)= H^k ( \partial) 
\end{equation}
and (\ref{eq:Max2}) is manifestly real. The Hamiltonians
(\ref{eq:Hphi}) and (\ref{eq:Hphilint3}) to be considered below both
satisfy (\ref{eq:ihom}), where in the case of (\ref{eq:Hphi}) the
correct branch of the square-root, $i^{4/2}=-1$, has to be chosen.

To support our claim that (\ref{eq:Max1}) and (\ref{eq:Max2}) are the correct 
Finsler versions of Maxwell's equations, we apply the operator $\partial _m$ 
to (\ref{eq:Max2}) for the case that $J_j=0$,
\begin{equation}\label{eq:wave0}
0 = \partial _m \big( H^k ( - i \partial ) F_{kj} \big) =
H^k ( - i \partial ) \big( \partial _m F_{kj} \big) \, .
\end{equation}
By (\ref{eq:Max1}), this can be rewritten as
\begin{gather}\label{eq:wave1}
0 = H^k ( - i \partial ) 
\big( \partial _k F_{jm} + \partial _j F_{mk} \big) 
\\
\nonumber
=
\partial _k \big( H^k ( - i \partial )  F_{jm} \big)+
\partial _j \big( H^k ( -i \partial ) F_{mk} \big) \, .
\end{gather}
The second term vanishes because of $J_m=0$. Using 
(\ref{eq:homH}) we find that $F_{jm}$ satisfies a 
generalized wave equation,
\begin{equation}\label{eq:wave}
H( - i \partial ) F_{jm} = 0 \, .
\end{equation}
If we solve this equation with a plane-wave ansatz for the 
electromagnetic field,
\begin{equation}\label{eq:pw}
F_{jm}(x) = \mathrm{Re} \Big\{f_{jm} \, \mathrm{exp}(ik_lx^l) \Big\} \, ,
\end{equation}
we find that the wave covector $k_l$ has to satisfy the equation
\begin{equation}\label{eq:eikonal}
H( k )  = 0 \, ,
\end{equation}
i.e., that in our flat Finsler space-time
electromagnetic waves propagate along lightlike straight lines.
This observation supports our claim that (\ref{eq:Max1})
and (\ref{eq:Max2}) are the correct Finsler versions of
Maxwell's equations.

To give further support to this claim, we now demonstrate that 
(\ref{eq:Max2}) can be brought into a form which 
is adapted to the formalism of premetric electrodynamics, 
cf.~\cite{birkbook}. To that end we have to show that
(\ref{eq:Max2}) can be rewritten as
\begin{equation}\label{eq:Max2d}
\partial_l {\mathcal{H}}{}^{ml} = - J^m \, ,
\end{equation}
where the excitation ${\mathcal{H}}{}^{ml}$ is related to the field 
strength $F_{kj}$ by a certain constitutive law. We write 
(\ref{eq:Max2}) in the equivalent form of (\ref{eq:Max2g}) 
and we apply the pseudo-differential operator $g^{mj} ( - i \partial )$. 
Then we obtain
\begin{equation}\label{eq:Max2b}
g^{mj}(-i \partial) g^{kl}(-i\partial) \partial_l F_{kj} = - \mu_0 J^m
\end{equation}
with $J^m = g^{mj}(-i\partial) J_j$. Since $g^{kl}$ is independent
of the $x^i$, this can be rewritten as
\begin{equation}\label{eq:Max2c}
\partial_l \left(\kappa^{mlkj}(- i \partial) F_{kj}\right) = - J^m
\end{equation}
with a constitutive operator
\begin{gather}\label{eq:kappa}
\kappa^{mlkj}(-i\partial) = 
\\
\nonumber
\frac{1}{2 \mu _0}
\Big(g^{mj}(-i\partial) g^{kl}(-i\partial) - 
g^{mk}(-i\partial) g^{jl}(-i\partial)\Big) \, .
\end{gather}
This form is equivalent to the original equation (\ref{eq:Max2}). In
particular, for $g^{ij}=\eta ^{ij}$ we return to the standard
Maxwell vacuum electrodynamics on Minkowski spacetime. We have, thus,
put our modified Maxwell equations in the premetric form, where the
constitutive law 
\begin{equation}\label{eq:const}
{\mathcal{H}}{}^{ml} = \kappa^{mlkj}(-i\partial) F_{kj} 
\end{equation}
involves the pseudo-differential operator (\ref{eq:kappa}).
An important advantage of the premetric formulation is that, quite
generally, (\ref{eq:Max2d}) together with the antisymmetry of 
$\mathcal{H}{}^{kl}$ immediately implies charge conservation,
$\partial _m J^m = 0$.

The homogeneous part of Maxwell's equations (\ref{eq:Max1}) is automatically 
satisfied if we express the 
electromagnetic field in terms of a potential,
\begin{equation}\label{eq:A}
F_{ij}  = \partial _i A_j - \partial _j A_i \, .
\end{equation}
We mention in passing that then the inhomogeneous part (\ref{eq:Max2c})
can be derived from the action
\begin{equation}
\mathcal{S} = 
\int \Big( \dfrac{1}{4}
\kappa^{klij}(- i \partial) F_{kl} (x) F_{ij} (x) 
- \mu _0 A_i (x) J^i (x) \Big)
d^4 x 
\end{equation}
where one has to take into account that the operator 
$\kappa ^{klij} ( -i \partial )$ commutes with the variational derivative.

In the following we will be interested in static fields. Then  
$\partial _0 A_i = 0$ and (\ref{eq:Max2}) implies
\begin{equation}\label{eq:HiA}
i H^k ( - i \partial ) \partial _k A_0  = - \mu _0 J_0 \, .
\end{equation}
We denote the four components of the potential by 
$(A_0=-V/c,A_1,A_2,A_3)$ and the four components of 
the current density by $(J_0=-c \rho, J_1,J_2,J_3)$. Then
(\ref{eq:HiA}) can be rewritten, with the help of (\ref{eq:homH}), as
\begin{equation}\label{eq:HA}
2 \, H ( - i \partial ) \,  V  = \dfrac{\rho}{\varepsilon _0} 
\end{equation}
where $\varepsilon _0$ is the permittivity of the vacuum and we have used
that $c^{-2} = \varepsilon _0 \mu _0$. If the metric is the unperturbed
Minkowski metric, we have of course $2 H( - i \partial ) V = - \triangle V$
where $\triangle$ is the ordinary Laplacian.
(\ref{eq:HA}) is the Finslerian
modification of the Poisson equation that determines the electrostatic
potential $V$ of a static charge density $\rho$. This is the only equation 
from Finslerian electrodynamics that we will need in the following.

\newpage

\section{The Finslerian modification of the Coulomb field}\label{eq:Coulomb}
\noindent
\subsection{A Finsler perturbation of Minkowski space--time}
We further specify our Finsler structure by assuming that the Hamiltonian (\ref{eq:H}) is a 
small perturbation of the standard Hamiltonian on Minkowski space--time. The latter reads
\begin{equation}\label{eq:H0}
H_0(p) = \, \dfrac{1}{2} \, \eta ^{ij}p_ip_j \, = \,
\dfrac{1}{2} \, \big( -p_0^2+\delta ^{\mu \nu}p_{\mu}p_{\nu} \big) \, .
\end{equation}
We restrict to the case that the 
Finsler perturbation affects the spatial part only. The simplest non-trivial ansatz for
such a perturbation is a square-root of a fourth-order term,
\begin{equation}\label{eq:Hphi}
H(p) = \, \dfrac{1}{2} \, \left( - p_0^2  + 
\sqrt{\big( \delta ^{\mu \nu} \delta ^{\rho \sigma}+ 4 \phi ^{\mu \nu \rho \sigma} \big)
p_{\mu}p_{\nu}p_{\rho}p_{\sigma} } \, \right)
\end{equation}
where $\phi ^{\mu \nu \rho \sigma}$ is totally symmetric. (A similar perturbation of
Minkowski spacetime was considered in \cite{Lammerzahl:2008di}.) We assume that the Finsler
perturbation is so small that we can linearize all equations with respect to the 
$\phi ^{\mu \nu \rho \sigma}$. Then the Hamiltonian simplifies to
\begin{equation}\label{eq:Hphilin}
H(p) = \dfrac{1}{2} \, \left( - p_0^2  + 
\delta ^{\rho \sigma} p_{\rho}p_{\sigma} 
\, + \, \dfrac{2 \phi ^{\mu \nu \rho \sigma} 
p_{\mu}p_{\nu}p_{\rho}p_{\sigma}
}{
\delta ^{\lambda \kappa}p_{\lambda} p_{\kappa}}
\, \right) .
\end{equation}
We will now demonstrate that the trace part of 
$\phi ^{\mu \nu \rho \sigma}$ can be 
eliminated with the help of a coordinate transformation. 
To that end, we decompose $\phi ^{\mu \nu \rho \sigma}$
in the form
\begin{equation}\label{eq:phitrace}
\phi ^{\mu \nu \rho \sigma} p_{\mu}p_{\nu}p_{\rho}p_{\sigma} \, = \, 
\big( \tilde{\phi}{}^{\mu \nu} \delta ^{\rho \sigma}  \, + \, 
\tilde{\phi}{}^{\mu \nu \rho \sigma} \big) p_{\mu}p_{\nu}p_{\rho}p_{\sigma} 
\end{equation}
where  $\tilde{\phi}{}^{\mu \nu \rho \sigma}$ is totally symmetric and 
trace-free.
Then (\ref{eq:Hphilin}) can be rewritten as
\begin{gather}\label{eq:Hphilint}
H(p) = \, 
\\
\nonumber
\dfrac{1}{2}  \Big( -  p_0^2 
+ 
\delta ^{\rho \sigma} p_{\rho} p_{\sigma} 
 + 2  \tilde{\phi}{}^{\rho \sigma} p_{\rho} p _{\sigma} 
 + 
\dfrac{2 \tilde{\phi}{}^{\mu \nu \rho \sigma} p_{\mu}p_{\nu}p_{\rho}p_{\sigma} \, 
}{
\delta ^{\lambda \kappa}p_{\lambda} p_{\kappa}}
 \Big)  .
\end{gather}
After a linear coordinate transformation,
\begin{equation}\label{eq:coordx}
\tilde{x}{}^0 = x^0 \, , \qquad 
\tilde{x}{}^{\sigma} \, = \, 
\big( \, \delta ^{\sigma}_{\mu} - \delta _{\mu \lambda} \tilde{\phi}{}^{\sigma \lambda} 
\, \big) x^{\mu} \, ,
\end{equation}
\begin{equation}\label{eq:coordp}
p_0 \, = \, \tilde{p}{}_0  \, , \qquad 
p_{\mu} \, = \, 
\big( \, \delta ^{\sigma}_{\mu} - \delta _{\mu \lambda} \tilde{\phi}{}^{\sigma \lambda} 
\, \big) \tilde{p}{}_{\sigma} \, ,
\end{equation}
the Hamiltonian reads
\begin{equation}\label{eq:Hphilint2}
H(\tilde{p}) = \, \dfrac{1}{2} \, \left( \, - \, \tilde{p}{}_0^2 
\, + \, 
\delta ^{\rho \sigma} \tilde{p}{}_{\rho} \tilde{p}{}_{\sigma}  \, + \, 
\dfrac{2 \tilde{\phi}{}^{\mu \nu \rho \sigma} 
\tilde{p}{}_{\mu}\tilde{p}{}_{\nu}\tilde{p}{}_{\rho}\tilde{p}{}_{\sigma} \, 
}{
\delta ^{\lambda \kappa}\tilde{p}{}_{\lambda} \tilde{p}{}_{\kappa}}
\, \right) 
\end{equation}
up to terms of quadratic order with respect to the Finsler perturbation.
If we drop the tilde, we have found the final form of our Hamiltonian,
\begin{equation}\label{eq:Hphilint3}
H(p) = \, \dfrac{1}{2} \, \left( \, \eta ^{ij} p_i p_j 
\, + \, 
\dfrac{2 \phi ^{\mu \nu \rho \sigma} 
p_{\mu} p_{\nu} p_{\rho} p_{\sigma} \, 
}{
\delta ^{\lambda \kappa} p_{\lambda} p_{\kappa}}
\, \right) \, ,
\end{equation}
with $\phi ^{\mu \nu \rho \sigma}$ totally symmetric and trace-free. A
totally symmetric fourth-rank tensor in three dimensions has 15 independent
components. The trace-free condition allows to express 6 of them in terms
of the other ones, e.g.
\begin{equation}\label{eq:trace}
\begin{split}
\phi^{1122}\, = \, \dfrac{1}{2} \, \big( \phi^{3333}-\phi^{1111}-\phi^{2222} \big) \, ,
\\
\phi^{1133}\, = \, \dfrac{1}{2} \, \big( \phi^{2222}-\phi^{3333}-\phi^{1111} \big) \, ,
\\
\phi^{2233}\, = \, \dfrac{1}{2} \, \big( \phi^{1111}-\phi^{2222}-\phi^{3333} \big) \, ,
\\
\phi^{1123}\, = \, - \phi^{2223}-\phi^{2333} \, , \qquad
\\
\phi^{1223}\, = \, - \phi^{1113}-\phi^{1333} \, , \qquad
\\
\phi^{1233}\, = \, - \phi^{1112}-\phi^{1222} \, , \qquad
\end{split}
\end{equation}
so we are left with 9 independent Finsler perturbation
coefficients. 

\subsection{The modified Coulomb field}
With the Hamiltonian (\ref{eq:Hphilint3}) inserted into (\ref{eq:HA}), we 
want to find the solution where the source is a point charge at rest. The 
equation we have to solve reads
\begin{equation}\label{fin-max8}
\triangle V + 2\,\frac{\phi^{\a\b\g\d}
\partial_\a\partial_\b\partial_\g\partial_\d}
{\triangle} V = \, - \, \dfrac{q}{\varepsilon _0} \, \delta( \vec{r})\,.
\end{equation}
Here and in the following we write
\begin{equation}\label{rDelta}
\vec{r} = (x^1,x^2,x^3) \, , \quad r = \sqrt{\delta_{\alpha \beta} x^{\alpha} x^{\beta}} \, , \quad
\triangle = \delta ^{\alpha \beta} \partial _{\alpha} \partial _{\beta} \,  .
\end{equation}
We look for a solution to (\ref{fin-max8}) in the form
\begin{equation}\label{fin-max9}
V ( \vec{r}) = \dfrac{q}{4 \pi \varepsilon _0 r}+\psi \big( \vec{r} \big) 
\end{equation}
where the first term on the right-hand side is the standard Coulomb solution of 
the unperturbed problem. As we agreed to linearize all equations with respect to 
the Finsler coefficients $\phi^{\alpha \beta \mu \nu}$, it is sufficient to 
determine $\psi$ to within this approximation. Then  $\psi$ must satisfy the equation
\begin{equation}\label{fin-max9x}
\triangle \psi+2\,\frac{\phi^{\a\b\g\d}
\partial_\a\partial_\b\partial_\g\partial_\d}
{\triangle} \left(\frac{q}{4 \pi \varepsilon _0 r} \right) = 0\,.
\end{equation}
Applying the Laplacian to this equation gives a linear fourth order PDE, 
\begin{equation}\label{fin-max10}
\triangle^2 \psi=-2 q \phi^{\a\b\g\d}
\partial_\a\partial_\b\partial_\g\partial_\d
\left(\frac{q}{4 \pi \varepsilon _0 r} \right) \, .
\end{equation}
The right-hand side of this equation is easily calculated, 
\begin{equation}\label{fin-max11}
\triangle^2 \psi= - \, \dfrac{210}{4 \pi \varepsilon _0}
\frac q{r^9} \, \phi^{\a\b\g\d}x_\a x_\b x_\g x_\d \, ,
\end{equation}
where $x_{\alpha} = \delta _{\alpha \beta} x ^{\beta}$. Here we have used 
that $\phi^{\alpha \beta \gamma \delta}$ is trace-free. 

The solution $\psi$ of the biharmonic equation (\ref{fin-max11}) must
\begin{itemize}
\item[(a)] be asymptotically zero for $r\to \infty$, 
\item[(b)] be linear with respect to $\phi^{\a\b\g\d}$,
\item[(c)] have only one singular point located at the origin, 
\item[(d)] be constructed from the $\phi^{\a\b\g\d}$ and the $x_\a$. 
\end{itemize}
Under these circumstances we can guess the solution of (\ref{fin-max11}) to be 
of the form
\begin{equation}\label{fin-max12}
\psi=C\frac{\phi^{\a\b\g\d}x_\a x_\b  x_\g x_\d}{r^5}\, .
\end{equation}
Note that we cannot add terms proportional to $\phi ^{\alpha \beta \gamma \delta}
\delta _{\alpha \beta} x_{\gamma} x_{\delta}$ or $\phi ^{\alpha \beta \gamma \delta}
\delta _{\alpha \beta} \delta _{\gamma \delta}$ because these terms vanish.
 
The biharmonic operator applied to (\ref{fin-max12}) gives 
\begin{equation}\label{res}
\triangle^2 \psi=
\frac{280C}{r^9} \, \phi^{\a\b\g\d}x_\a x_\b x_\g x_\d \, . 
\end{equation}
By comparing (\ref{res}) with (\ref{fin-max11}) we obtain $C=- 3 q (16 \pi \varepsilon _0)^{-1} $. 
Thus the solution of (\ref{fin-max10}) is
\begin{equation}
\psi=-\frac{3 q}{16 \pi \varepsilon _0 r^5} \,  
\phi^{\a\b\g\d}x_\a x_\b  x_\g x_\d \, .
\end{equation}
Consequently, we have the scalar potential of the point source in the form
\begin{equation}
V=\frac{q}{4 \pi \varepsilon _0 r} 
\left(1-\frac{3}{4r^4} {\phi^{\a\b\g\d}x_\a x_\b  x_\g x_\d}\right) \,.
\end{equation}
In spherical coordinates this expression reads
\begin{equation}\label{Coul}
V=\frac{q}{4 \pi \varepsilon _0 r} 
\left(1-\frac{3}{4} {\phi^{\a\b\g\d}f_{\a\b\g\d}(\theta, \varphi) }\right) 
\end{equation}
where 
\begin{gather}
\nonumber
\phi^{\a\b\g\d}f_{\a\b\g\d}(\theta, \varphi) 
=
\phi ^{1111} \mathrm{sin} ^4 \theta
\, \mathrm{cos} ^4 \varphi 
\\
\nonumber
+ 
\phi^{1112} \mathrm{sin} ^4 \theta \, \mathrm{cos} ^3 \varphi
\, \mathrm{sin} \, \varphi 
+ 
\phi^{1113} \mathrm{sin} ^3 \theta \,
\mathrm{cos} \, \theta \,  \mathrm{cos} ^3 \varphi
\\
\nonumber
+ 
\phi^{1122} \mathrm{sin} ^4 \theta
\, \mathrm{cos} ^2 \varphi \, \mathrm{sin} ^2 \varphi
+ 
\phi^{1123} \mathrm{sin} ^3 \theta \, 
\mathrm{cos} \, \theta \, \mathrm{cos} ^2 \varphi\, 
\mathrm{sin} \, \varphi
\\
\nonumber
+ 
\phi^{1133} \mathrm{sin} ^2 \theta \, 
\mathrm{cos} ^2 \theta \, \mathrm{cos} ^2 \varphi
+ 
\phi^{1222} \mathrm{sin} ^4 \theta \,
\mathrm{cos} \, \varphi \, \mathrm{sin} ^3 \varphi
\\
\nonumber
+ 
\phi^{1223} \mathrm{sin} ^3 \theta \,
\mathrm{cos} \, \theta \, \mathrm{cos} \, \varphi \,
\mathrm{sin} ^2 \varphi
\\
\nonumber
+ 
\phi^{1233} \mathrm{sin} ^2 \theta \,
\mathrm{cos} ^2 \theta \,  \mathrm{cos} \, \varphi
\, \mathrm{sin} \, \varphi
+ 
\phi ^{1333} \mathrm{sin} \, \theta \,
\mathrm{cos} ^3 \theta \, \mathrm{cos} \, \varphi
\\
\nonumber
+ 
\phi^{2222} \mathrm{sin} ^4 \theta \,
\mathrm{sin} ^4 \varphi 
+ 
\phi^{2223} \mathrm{sin} ^3 \theta
\, \mathrm{cos} \, \theta \,
\mathrm{sin} ^3 \varphi
\\
\nonumber
+ 
\phi^{2233}
\mathrm{sin} ^2 \theta \, 
\mathrm{cos} ^2 \theta \, \mathrm{sin} ^2 \varphi
+ 
\phi^{2333} \mathrm{sin} \, \theta \,
\mathrm{cos} ^3 \theta \, 
\mathrm{sin} \, \varphi
\\
+ 
\phi^{3333} \mathrm{cos} ^4 \theta \; .
\label{eq:phif}
\end{gather}


\section{Finsler corrections of the hydrogen energy levels}\label{eq:shift}
\subsection{Finsler modified Schr{\"o}dinger equation}
For an electron (mass$\, =m$ and charge$\, =-e$) in the Coulomb field (\ref{Coul}) of a proton
(charge $q=e$), the Schr{\"o}dinger equation reads
\begin{gather}\label{Schr}
-\frac{\hbar^2}{2m}
\left( \triangle + 2 \, 
\dfrac{\phi ^{\alpha \beta \gamma \delta} 
\partial _{\alpha} \partial _{\beta} \partial _{\gamma}\partial _{\delta}}{\triangle}
\right) \Psi \big( \vec{r} \, \big) 
\\
-\frac{e^2}{4 \pi \varepsilon _0 r} 
\left( \, 1 \, - \, \frac{3}{4} \phi^{\a\b\g\d}f_{\a\b\g\d}(\theta, \varphi)
\right) 
\Psi \big( \vec{r} \big)  = E \Psi \big( \vec{r} \, \big) \, .
\nonumber
\end{gather}
Here we have added to the potential term a Finsler correction according to our
results from the preceding section, and we have added to the Laplacian the same 
correction as in the electrodynamic equations, cf. (\ref{fin-max8}). 
The latter assumption is based on the 
idea that the Finsler perturbation modifies the underlying geometry such
that particles and light are affected in the same way. As an alternative, one
might speculate that there are two different Finsler modifications of the 
space-time structure, one for particles and one for light. This would come up
to a Finslerian bimetric theory. We will not investigate such a more complicated
theory here but rather stick with (\ref{Schr}). However, we mention that
the order-of-magnitude estimates of the following calculations remain true
for the more general (bimetric) theories as long as the perturbation of the
Laplacian term does not exceed the corresponding term in (\ref{Schr}) by several
orders of magnitude.

To give further support to our Schr{\"o}dinger equation (\ref{Schr}), we
demonstrate that it comes about as the non-relativistic limit of a modified 
Klein-Gordon equation. The free Klein-Gordon equation in a Finsler space-time 
is naturally given by
\begin{equation}\label{eq:KG1}
2 H(- i \hbar \partial) \Phi + m^2 c^2 \Phi = 0
\end{equation}
where $H$ is the 4-dimensional Hamiltonian. This can also be derived 
from an action principle. In our model,
\begin{gather}\label{eq:KG2}
2 H(- i \hbar \partial) = 
\\
\nonumber
\hbar ^2 \left(
\frac{1}{c^2} \partial_t^2 -
\delta^{\mu\nu} \partial_\mu \partial_\nu 
-
\dfrac{2 \phi^{\alpha \beta \gamma \delta}
\partial_{\alpha} \partial_{\beta} 
\partial_{\gamma} \partial_{\delta}
}{
\delta^{\tau \lambda} \partial_{\tau} \partial_{\lambda} 
}
\right) 
\, .
\end{gather}
We want to derive the non-relativistic limit of this Finslerian Klein-Gordon equation. For 
that we use the formalism described in \cite{KieferSingh}. We make an ansatz where the 
wave function is given by an exponential function of a sum of terms of different orders 
of $c^{-2}$,
\begin{equation}\label{eq:KS0}
\Phi (x) = \exp\left(\frac{i}{\hbar}
\left(c^2 S_0(x) + S_1(x) + c^{-2}S_2(x) + \ldots\right)\right) \, .
\end{equation}
Here the functions $S_N (x)$ may take complex values.
As we are looking for solutions to (\ref{eq:KG1}) that are 
small perturbations of plane harmonic waves $\sim e^{i k_i x^i}$
and, hence, have no zeros, the ansatz (\ref{eq:KS0}) is no restriction of
generality. We insert this ansatz into the Klein-Gordon equation and 
equate equal powers of $c$. The 
equation of leading order, $c^4$, is
\begin{equation}\label{eq:KS1}
\big( \delta ^{\mu \nu} \delta ^{\rho \sigma}
+ 2 \phi ^{\mu \nu \rho \sigma} \big)
\partial _{\mu} S_0 \partial _{\nu} S_0 
\partial _{\rho} S_0 \partial _{\sigma} S_0 
= 0 \, .
\end{equation}
As the Finsler coefficients are small, this implies
$\partial _{\mu} S_0 = 0$, i.e., $S_0$ can only be a 
function of time, $S_0(x) = S_0(t)$. The next order, 
$c^2$, yields the equation
\begin{equation}\label{eq:KS2}
\left(\frac{dS_0(t)}{dt}\right)^2 - m^2 = 0 \, ,
\end{equation}
which possesses the solutions
\begin{equation}\label{eq:KS3}
S_0(t) = \pm \, m \, t +\mathrm{constant} 
\end{equation}
where, for physical reasons, we do not consider the plus sign. 
The equation of next order, $c^0$, gives for the function 
$\Phi _1(x) = e^{\frac{i}{\hbar} S_1(x)}$ the equation of motion
\begin{equation}
i \hbar \frac{\partial \Phi _1(x)}{\partial t} = 
- \frac{\hbar^2}{2m} \Big(  \triangle +
\dfrac{2 \phi^{\alpha \beta \gamma \delta}
\partial_{\alpha} \partial_{\beta} 
\partial_{\gamma} \partial_{\delta}
}{ \triangle} \Big) \Phi _1 (x) \, .
\end{equation}\label{eq:FiSch1}
This represents the free
Schr\"odinger equation in our Finsler space-time. Coupling 
to an electrostatic potential $V$ will be performed through 
\begin{equation}\label{eq:minimal}
\frac{\partial}{\partial t} \mapsto 
\frac{\partial}{\partial t} 
- \frac{i}{\hbar} \, e \, V (x) 
\end{equation}
which gives us the time-dependent Schr{\"o}dinger equation
with coupling to an electrostatic potential, 
\begin{gather}\label{eq:FiSch2}
i \hbar \frac{\partial \Phi _1(x)}{\partial t} = 
\\
\nonumber
- \frac{\hbar^2}{2m} \Big(  \triangle +
\dfrac{2 \phi^{\alpha \beta \gamma \delta}
\partial_{\alpha} \partial_{\beta} 
\partial_{\gamma} \partial_{\delta}
}{ \triangle} \Big) \Phi _1 (x)
- e \, V (x) \Phi _1(x) \, .
\end{gather}
Upon inserting for $V$ our expression for the perturbed Coulomb potential,
the time-independent Schr{\"o}dinger equation (\ref{Schr}) results from a 
separation ansatz $\Phi _1 (x) = \Psi \big( \vec{r} \big) e^{-i Et/\hbar}$.
Note that in (\ref{Schr}) the radial variable $r$ can be separated from 
the angular variables $\theta$ and $\vp$ exactly as in the ordinary theory.  
The two angular variables, however, cannot be separated from each other.
 
 
\subsection{Finsler modified energy levels}
We want to determine the bound states and the energy levels by the perturbation 
method to within linear order in the Finsler coefficients $\phi^{\a\b\g\d}$. This
will give us the splitting of the hydrogen spectral lines as produced by the
Finsler perturbation. Of course, as we are considering the simple Kepler problem
as the unperturbed situation, this splitting is to be viewed on top of all the
other (fine-structure and hyperfine-structure) splittings of the hydrogen spectral 
lines which are well understood. 

We denote the unperturbed bound states of the Coulomb potential by
\begin{gather}\label{upsi}
\Psi_{n l m} \big( \vec{r} \, \big) =  
\\{\nonumber}
\sqrt{ \dfrac{2^3 (n - l - 1)!}{n^3 a_0^3 2 n (n + l)!}} \, 
e^{-\frac{r}{n a_0}} \left( \dfrac{2 r }{n a_0} \right)^l 
L_{n - l - 1}^{2 l + 1} \Big( \dfrac{2 r}{n a_0} \Big) 
  \, Y_l^m(\theta, \vp ) 
\end{gather}
where 
\begin{equation}\label{a0}
a_0 \, = \, \dfrac{4\pi \varepsilon _0 \hbar^2}{me^2}
\end{equation}
is the Bohr radius, the $L_p^q$ are the generalized Laguerre polynomials and 
the $Y_l^m$ are the spherical harmonics. The quantum numbers $n, l$ and $m$ 
take the values 
\begin{equation}\label{numb}
n=1,2,\cdots\,;\quad l=0,\cdots\!, n-1; \quad m=0,\cdots\!, \pm l.
\end{equation}
The corresponding unperturbed eigenvalues are
\begin{equation}\label{En}
E_n \, = \, - \, \dfrac{\mathrm{Ry}}{n^2} \, , \quad
\mathrm{Ry} \, = \, \dfrac{e^2}{8 \pi \varepsilon _0 a_0} \, .
\end{equation}
The first-order corrections to the eigenvalues are determined by the
matrix elements 
\begin{gather}\label{M}
M_{nlm,n'l'm'} =
- \, 
\left< \Psi_{nlm}\, \Big| \, 
\frac{ \hbar ^2\phi^{\a\b\g\d} 
\partial _{\a} \partial _{\b} \partial _{\g} \partial _{\d}}{m \, \triangle} \, 
\Psi_{n'l'm'} \right> 
\nonumber
\\
+ \, \left< \Psi_{nlm}\, \Big| \, 
\frac{3e^2}{16 \pi \varepsilon _0 r} \phi^{\a\b\g\d}f_{\a\b\g\d}(\theta, \varphi) \, 
\Psi_{n'l'm'} \right> \, .
\end{gather}
The first scalar product on the right-hand side can be calculated
more easily in the momentum representation,
\begin{gather}\label{Mp}
- \,  
\left< \Psi_{nlm}\, \Big| \, 
\frac{ \hbar ^2 \phi^{\a\b\g\d} 
\partial _{\a} \partial _{\b} \partial _{\g} \partial _{\d}}{m \, \triangle} \, 
\Psi_{n'l'm'} \right> 
\\
= \,
\nonumber
\dfrac{1}{m} \, 
\left< \hat{\Psi}{}_{nlm}\, \Big| \, 
 \phi^{\a\b\g\d} f_{\a\b\g\d}(\theta , \varphi ) p^2 
\hat{\Psi}{}_{n'l'm'} \right>
\end{gather}
where $\hat{\Psi}{}_{nlm} \big( \vec{p} \, \big)$ is the Fourier transform
of $\Psi _{nlm} \big( \vec{r} \, \big)$ which is given by \cite{PodolskyPauling}
\begin{gather}\label{ppsi}
\hat{\Psi}{}_{n l m} \big( \vec{p} \, \big) =  
\sqrt{\dfrac{2 a_0^3 \hbar n(n-l-1)!}{\pi (n+l)!}}
\\{\nonumber}
\times \dfrac{2^{2l+2}(\hbar a_0 p)^l}{(a_0^2 p^2+\hbar ^2 )^{l+2}}
\, C^{l+1}_{n-l-1} 
\Big( \dfrac{a_0^2 p^2 - \hbar ^2}{a_0^2 p^2 + \hbar ^2} \Big)
\, Y_l^m(\theta, \vp ) 
\end{gather}
where the $C^k_s$ are the Gegenbauer polynomials. 

We now calculate the necessary matrix elements one by one to determine
the perturbations of the lowest energy levels.
 
The ground state, $n=1$, is non-degenerate. Under the Finsler perturbation, its 
energy value is shifted in first-order perturbation theory according to 
\begin{equation}\label{E1}
E_1 \to E_1 + \Delta E_1
\end{equation}
where
\begin{equation}\label{DE1}
\Delta E_1 = M_{100,100}
 \, .
\end{equation}
Calculation of this matrix element yields
\begin{equation}\label{DE1n}
\Delta E_1  =  \dfrac{7 \, \mathrm{Ry}}{12} \, 
\big(  \phi^{1111} + \phi^{2222} + \phi^{3333} \big)
\end{equation}
where we have used the trace-free condition.

The next level, $n=2$, is fourfold degenerate in the unperturbed 
situation. Under the Finsler perturbation, it will in general split 
into four levels,
\begin{equation}\label{E2}
E_2 \to E_2 + \Delta E_2^A \, , \qquad A=1,2,3,4
\end{equation}
where, in first-order perturbation theory, the $\Delta E_2^A$ are
the eigenvalues of the perturbation matrix $\big( M_{2lm,2l'm'} \big)$.
The entries of this $(4 \times 4)-$matrix can be calculated. Using
again the trace-free condition, we find
\begin{equation}\label{M1}
M_{200,200}  =   \dfrac{19 \, \mathrm{Ry}}{48} \, 
\big(  \phi^{1111} + \phi^{2222} + \phi^{3333} \big) \, ,
\end{equation}
\begin{equation}\label{M2}
M_{210,210}  =  
\dfrac{19 \, \mathrm{Ry}}{112} \, \Big(
\phi^{1111} + \phi^{2222} + 5 \phi^{3333} \Big) \, ,
\end{equation}
\begin{gather}\label{M3}
M_{211,211}  =  M_{21(-1),21(-1)} = 
\\ 
 =   
\dfrac{19 \, \mathrm{Ry}}{112} \, \big(
3 \phi^{1111} + 3 \phi^{2222} + \phi^{3333} \big)
\, , \nonumber
\end{gather}
\begin{equation}\label{M4}
M_{200,210} = M_{200,211} = M_{200,21(-1)} = 0  \, ,
\end{equation}
\begin{gather}\label{M5} 
M_{210,211} = - \overline{M_{210,21(-1)}} = \\
 = \, - \, 
\dfrac{19 \, \mathrm{Ry}}{\sqrt{2} \, 140} \,
\Big( \phi^{1113} + \phi^{1333} 
+ i \,  \big( \phi^{2223} + \phi^{2333} \big) \Big)
\, ,\nonumber
\end{gather}
\begin{gather}\label{M6}
M_{211,21(-1)} = 
\\
= \dfrac{19 \, \mathrm{Ry}}{56} \,
\Big( - \phi^{1111} +  \phi^{2222} 
+ \dfrac{2i}{5} \, \big( \phi^{1112} + \phi^{1222} \big) \Big)
\, , \nonumber
\end{gather}
where overlining means complex conjugation. 

The perturbation matrix consists of a $1 \times 1$ block and a $3 \times 3$ block.
Therefore, calculating the eigenvalues requires solving a third-order equation. This
can be done explicitly, but the resulting expressions are rather awkward and will 
not be given here.

The transition from the $E_2$ level to the $E_1$ level is known as the Lyman-$\alpha$ 
line. Our Finsler perturbation causes a splitting of this line into four lines in
general, a singlet ($l=0$) and a triplet ($l=1$). The Lyman-$\alpha$ line does
\emph{not} split if and only if the perturbation matrix  $M_{2lm,2l'm'}$ is a multiple
of the unit matrix. This is the case if and only if
\begin{gather}\label{eq:la}
\phi ^{1111} =\phi ^{2222} = \phi ^{3333} = 0
\\
\nonumber
\phi ^{1112}+ \phi ^{1222} = \phi ^{1113}+ \phi ^{1333} = \phi ^{2223}+ \phi ^{2333} = 0
\, .
\end{gather}
(The six Finsler coefficients on the left-hand sides of (\ref{eq:trace}) 
are then all zero.) This demonstrates that observations of the 
Lyman-$\alpha$ line alone cannot give us bounds on \emph{all} Finsler 
coefficients. Even if we observe, with a certain measuring accuracy, 
that the Lyman-$\alpha$ line does not split, we could have arbitrary 
Finsler coefficients $\phi ^{1112}=- \phi ^{1222}$, 
$\phi ^{1113}=- \phi ^{1333}$ and $\phi ^{2223}=
- \phi ^{2333}$.

One may consider the transition from the $E_3$ level to the $E_1$ level in 
addition which, in the unperturbed situation, gives rise to the Lyman-$\beta$ line.
The Lyman-$\beta$ line splits, in general, into nine lines, a singlet ($l=0$),
a triplet ($l=1$) and a quintuplet ($l=2$). The energy shifts are determined by 
the eigenvalues of the matrix  $(M_{3lm,3l'm'})$. We calculate only two of these matrix
elements, 
\begin{gather}\label{M37}
M_{322,32(-2)} =
\\
\nonumber
\dfrac{13 \, \mathrm{Ry}}{252} \, \Big(
3 \, \phi^{1111} + 3 \, \phi^{2222} - \phi^{3333} + 2 \, i \big( \phi^{1222}-\phi^{1112} \big) \Big) 
\end{gather}
and 
\begin{equation}\label{M38}
M_{320,321}  = - \overline{M_{320,32(-1)}} =
- \, \dfrac{13 \, \mathrm{Ry}}{\sqrt{6} \, 42} \, 
\Big( \phi^{1333} + i \, \phi^{2333} \Big) 
\, . 
\end{equation}
The Lyman-$\beta$ line does not split if and only
if the matrix $(M_{3lm,3l'm'})$ is a multiple of the unit matrix.
This requires, in particular, vanishing of the two off-diagonal 
matrix elements we have calculated, hence
\begin{equation}\label{eq:lb}
\phi^{1222}-\phi^{1112} =  \phi^{1333} = \phi^{2333} = 0 \; .
\end{equation}
If neither the Lyman-$\alpha$ nor the Lyman-$\beta$ line splits, both
(\ref{eq:la}) and (\ref{eq:lb}) have to hold, so in this case all 
Finsler coefficients must be zero. This demonstrates that we get 
bounds on \emph{all} Finsler coefficients if we observe, with a certain
measuring accuracy, that neither the Lyman-$\alpha$ line nor the 
Lyman-$\beta$ line splits.  

As a special case, we consider a Finsler perturbation that 
respects the symmetry about the $z$-axis. This simplifying
assumption seems reasonable in a laboratory on Earth if one believes 
that the Finsler anisotropy has a gravitational origin. Then the 
expression $\phi^{\a\b\g\d}f_{\a\b\g\d}(\theta, \varphi)$ in 
(\ref{Coul}) must be independent of $\varphi$. In combination with 
the trace-free condition, this symmetry assumption requires 
that (\ref{eq:phif}) simplifies to
\begin{equation}\label{eq:rotsym}
\phi^{\a\b\g\d}f_{\a\b\g\d}(\theta, \varphi) \, = \, 
\phi^{1111} \big( 1-5 \, \mathrm{cos} ^2 \theta + 10 \, 
\mathrm{cos} ^4 \theta \big)
\, ,
\end{equation}
i.e., there is only one independent Finsler coefficient left. 

The perturbation of the $E_1$ level (\ref{DE1n}) simplifies to
\begin{equation}\label{DE1nsym}
\Delta E_1  =  
\dfrac{14 \, \mathrm{Ry}}{3} \,  \phi^{1111} \, .
\end{equation}
The perturbation matrix  $(M_{2lm,2l'm'})$ becomes diagonal, so that the eigenvalues
can be easily calculated. For the singlet we find
\begin{equation}\label{DE20sym}
\Delta E_2^1  =  
\dfrac{19 \, \mathrm{Ry}}{6} \,  \phi^{1111} \, ,
\end{equation}
whereas the triplet degenerates into two lines,
\begin{equation}\label{DE212sym}
\Delta E_2^2  =  
\dfrac{38 \, \mathrm{Ry}}{7} \,  \phi^{1111} 
\, , 
\end{equation}
\begin{equation}\label{DE213sym}
\Delta E_2^3  =  \Delta E_2^4  =
\dfrac{57 \, \mathrm{Ry}}{28} \,  \phi^{1111} \, .
\end{equation}
This demonstrates that, in this case, the Lyman-$\alpha$ line splits 
into three lines. The spacing between the outermost lines is
\begin{equation}\label{split}
\Delta E_2^2  - \Delta E_2^3  =
\dfrac{95 \, \mathrm{Ry}}{28} \,  \phi^{1111} \; .
\end{equation}
If we observe, with a certain measuring accuracy $\delta \omega$ of
the frequency, that the Lyman-$\alpha$ line does not split, we can
deduce that   
\begin{equation}\label{deltaE}
\big| \phi ^{1111} \big|  \, \le \,  
\dfrac{28 \,  \hbar \, \delta \omega}{95 \, \mathrm{Ry}}  
\approx 1.4 \times 10^{-17} \, \delta \omega / \mathrm{Hz}.
\end{equation}
In the general case, without the special symmetry assumption, we
get similar bounds for all Finsler
coefficients from the observation that neither the Lyman-$\alpha$
nor the Lyman-$\beta$ line splits. (Instead of $28/95$, we get
of course other numerical factors.)

\section{Conclusions}
We have calculated the Finsler perturbation of atomic
spectra for the simplest possible case, using the 
Schr{\"o}dinger equation with the standard Coulomb 
potential for the unperturbed situation and a linearized
metric perturbation that derives from the square-root
of a fourth-order term. We emphasize again that,
if the results are to be compared with measurements 
of the hydrogen spectrum, the Finslerian splitting of 
the spectral lines has, of course, to be viewed as 
coming on top of all the other fine-structure and 
hyperfine-structure splittings that
are well understood. Also, more complicated atomic
spectra and more complicated Finslerian metric 
perturbations can be considered. What we wanted to
estimate was the order of magnitude for the 
bounds on the Finsler perturbations that can be 
achieved by atomic spectroscopy. We see from 
(\ref{deltaE}) that these bounds are quite tight.
Given the fact that, nowadays, frequencies can be
measured in the optical and in the ultraviolet with
an accuracy of up to $\delta \omega \approx 
10^{-7}\;{\rm Hz}$, with this kind of measurements
it should be possible to get an upper bound on
the dimensionless Finsler coefficients of about 
$10^{-24}$. This bound is by several orders of magnitude
smaller than the bounds from Solar system tests, 
cf.  \cite{LaemmerzahlPerlickHasse2012}.

Using nuclear spectroscopy, rather than atomic 
spectroscopy, it might be possible to get even better 
bounds. The Hughes-Drever experiment
(see, e.g. Will \cite{Will}) comes to mind which
gives the best bounds on anisotropic mass terms 
to date. It is based on magnetic resonance measurements 
of a Li-7 nucleus whose ground state of spin 3/2 splits
into four levels when a magnetic field is applied.
Anisotropic mass terms would lead to an unequal 
spacing between these levels.
It was also shown that the Hughes-Drever 
experiment gives very restrictive bounds on torsion, see
\cite{Lammerzahl:1997}. The Finsler perturbations 
discussed in this paper are not exactly of the same
mathematical form as anisotropic mass terms or 
torsion terms, but they also introduce some kind
of spatial anisotropy. For this reason, it seems 
likely that a careful re-analysis of the Hughes-Drever
experiment would also give some strong bounds on 
possible Finsler perturbations, probably even stronger
than the bounds from atomic spectroscopy. However, there
are two difficulties with the Hughes-Drever experiment,
one from the theoretical side and one from the 
experimental side. Theoretically, the analysis of
the experiment would have to be based on a wave 
equation for a particle with spin, i.e., on a 
Finsler generalisation of a Dirac-type equation
or on a non-relativistic approximation thereof.
The basic idea of how such a Dirac-type equation 
could be found in a Finsler setting is rather 
straight-forward: One would have to linearize
the corresponding Klein-Gordon equation with 
respect to the derivative operators, see e.g.  
\cite{AudretschLammerzahl:1993}. However, the
procedure is considerably more complicated than
in the spinless case and the details have not
yet been worked out for the kind of Finsler
perturbation discussed in this paper. Experimentally,
a Hughes-Drever experiment in its standard setting
is performed by keeping the magnetic field fixed 
in the laboratory and waiting for 24 hours so 
that the Earth makes a full rotation with respect
to the spacetime background geometry. In this 
way, one can detect ``cosmological'' anisotropies, 
i.e, anisotropies in the background geometry,
but not ``gravitational'' anisotropies which 
would rotate with the Earth. If one thinks of
a Finsler perturbation as having a gravitational
origin, it would be of a type that could not be
detected with a Hughes-Drever experiment in its
usual setting. One would have to rotate the 
magnetic field with respect to the laboratory
which is technically more difficult.

For these two reasons, we have restricted in
this paper to a test with atomic spectroscopy,
rather than with nuclear spectroscopy of the
Hughes-Drever type. It should be noted that
such an atomic spectroscopy test applies not
only to laboratory experiments on Earth, but 
to any situation where (hydrogen) spectral 
lines are observed. So it can be used also for 
estimating Finsler perturbations in the 
neighborhood of distant stars or gas clouds. 

\section*{Acknowledgments}
We wish to thank the German-Israeli Foundation (GIF) for
supporting this work within the project ``Exploring the 
full range of electrodynamics''. During part of this
work, VP was financially supported by Deutsche 
Forschungsgemeinschaft, Grant LA905/10-1. Also,
CL and VP gratefully acknowledge support from 
the Deutsche Forschungsgemeinschaft within the 
Research Training Group 1620 ``Models of Gravity''. 
Moreover, we thank Nico Giulini for an important comment.


\end{document}